\documentclass[twocolumn,showpacs,preprintnumbers,amsmath,amssymb,superscriptaddress]{revtex4}

\usepackage{graphicx}
\usepackage{dcolumn}
\usepackage{bm}

\begin{document}


\title{Experimental observation of anisotropic diffraction induced by orbital angular momentum}

\author{Guo Liang}
\email[The first two authors have equally contributed to this work, with the first one mainly to the theory and the second one mainly to the experiments.]{}
\affiliation{Guangdong Provincial Key Laboratory of Nanophotonic Functional Materials and Devices, South China Normal University, Guangzhou 510631, China}
\affiliation{School of Physics and Electrical Information, Shangqiu Normal University, Shangqiu 476000, China}
\author{Huicong Zhang}
\email[The first two authors have equally contributed to this work, with the first one mainly to the theory and the second one mainly to the experiments.]{}
\affiliation{Guangdong Provincial Key Laboratory of Nanophotonic Functional Materials and Devices, South China Normal University, Guangzhou 510631, China}
\author{Lei Fang}
\affiliation{Guangdong Provincial Key Laboratory of Nanophotonic Functional Materials and Devices, South China Normal University, Guangzhou 510631, China}
\author{Qian Shou}
\affiliation{Guangdong Provincial Key Laboratory of Nanophotonic Functional Materials and Devices, South China Normal University, Guangzhou 510631, China}
\author{Wei Hu}
\affiliation{Guangdong Provincial Key Laboratory of Nanophotonic Functional Materials and Devices, South China Normal University, Guangzhou 510631, China}
\author{Qi Guo}
\email[Corresponding author's email address:]{guoq@scnu.edu.cn}
\affiliation{Guangdong Provincial Key Laboratory of Nanophotonic Functional Materials and Devices, South China Normal University, Guangzhou 510631, China}

\date{\today}

\begin{abstract}
We have predicted [Opt. Express, \textbf{26}, 8084 (2018)] that an orbital angular momentum (OAM) carried by the elliptic Gaussian beam (EGB) with a cross-phase
could induce an anisotropic diffraction (AD). When the OAM was equal to a specific value called a critical OAM, such an OAM-induced AD would make the spiraling elliptic Gaussian mode (fundamental eigenmode) exist in linearly isotropic media, where only the circular Gaussian mode is supposed to exist for the beam without the OAM. Here we experimentally demonstrate such a novel phenomenon via the propagations of the EGBs with the OAM in linearly and both-linearly-and-nonlinearly isotropic media, respectively. In the former case, when its carrying OAM equals the critical OAM, the spiraling elliptic Gaussian mode is observed in the free space. In the latter case, when the OAM and the power equal respectively the critical OAM and the critical power, the spiraling elliptic soliton, predicted by Desyatnikov et al. [Phys. Rev. Lett, \textbf{104}, 053902 (2010)], is observed for the first time to stably propagate in the cylindrical lead glass. The power-controllable rotation of such an EGB is also experimentally demonstrated, which may have applications to optical spanners in bio-photonics, opto-fluidics, and life sciences.
\end{abstract}

\maketitle

The diffraction phenomenon of light, the linear and ubiquitous effect of the light propagation, has been discovered as early as 15th century~\cite{Born-book}.
Due to the diffraction, the evolution of the (1+2)-dimensional paraxial optical beam, $\Psi$, is described by the paraxial equation~\cite{Fleck-josa-1983,Ciattoni-pre-2002,Chen-OC-2011,Guo-JOA-2000} in a dimensionless coordinate system~\cite{Liang-OE-2018}
$i{\partial_\zeta\Psi}+({1}/{2})\left(\delta_{\xi\xi}{\partial^2_{\xi\xi}\Psi}+\delta_{\eta\eta}{\partial^2_{\eta\eta}\Psi}\right)=0$.
That $\delta_{\xi\xi}\neq \delta_{\eta\eta}$ is general case, and is called the anisotropic diffraction (AD)~\cite{Polyakov-PRE-2002,Liang-OE-2018}. Otherwise, that $\delta_{\xi\xi}= \delta_{\eta\eta}(=1)$ is special case for the ordinary light in the uniaxial crystal or light in the isotropic medium, and the diffraction is isotropic.
Thus, the fundamental mode (Gaussian mode) among all eigenmodes of the paraxial equation must be transversely circular in the isotropic case~\cite{Kogelnik-AO-1966,Haus-1984} and asymmetrical in the anisotropic case~\cite{Liang-OE-2018}.
As a result, only the circular Gaussian beam and the elliptic Gaussian beam (EGB)
can expand their cross-sectional shapes uniformly (without changing their ellipticity) during the propagation in the (linear) isotropic~\cite{Liang-OE-2018,Kogelnik-AO-1966,Haus-1984} and anisotropic~\cite{Liang-OE-2018,Seshadri-JOSAA-2001,Seshadri-JOSAA-2003,Ciattoni-oc-2004} cases, respectively. Moreover, spatial solitons can be formed because of an exact balance between the diffraction effect and the nonlinear effect. Thus, only the fundamental soliton with cylindrical-symmetry
 can exist in the medium with both linear and nonlinear isotropy. The EGBs in such a medium, however, always undergo significant oscillations~\cite{Crosignani-OL-1993,Snyder-OL-1997,Tichonenko-OL-1998}, and generally cannot form elliptic spatial solitons~\cite{Eugenieva-OL-2000,Katz-OL-2004}. To obtain coherent elliptic solitons, either linear anisotropy~\cite{Conti-PRE-2005,Guo-JOA-2000} or nonlinear anisotropy~\cite{Zhang-OE-2007,Rotschild-PRL-2005}, or both~\cite{Polyakov-PRE-2002} should be introduced, while incoherent elliptic solitons came true by means of the anisotropy of the transverse correlation function~\cite{Eugenieva-OL-2000,Katz-OL-2004,Christodoulides-PRL-1998}. The coherent elliptic soliton could survive in nematic liquid crystals
with birefringence (linear anisotropy)~\cite{Conti-PRE-2005}, and was observed in the lead glass~\cite{Rotschild-PRL-2005}, where the nonlinear anisotropy is achieved by rectangular boundaries.

Different from a long history of the diffraction, it was 1992 when Allen and his coworkers~\cite{Allen-pra-1992} recognised that optical beams can carry orbital angular momenta (OAM), which 
have drawn growing interest owing to their potential applications, such as optical manipulations~\cite{Grier-Nature-2003,Dholakia-NP-2011}, and so on. Such optical beams are usually associated with optical vortices and related donut-shaped beams, such as the Laguerre-Gauss beams~\cite{Allen-pra-1992}, the Bessel beams~\cite{Volke-Sepulveda-job-2002}, and the other hollow beams~\cite{Gutierrez-Vega-proc-spie-2008}. There is, however, another kind of vortex-free beam that can also carry the OAM, which is an EGB with the cross-phase term~\cite{Desyatnikov-PRL-2010,Liang-PRA-2013,Allen-book-1999}
and is also called as an astigmatic beam~\cite{Courtial-OC-1997, Kotlyar-OE-2018}.

Although the diffraction and the OAM, as two essential and intrinsic properties of optical beams,
were found to have very rich interaction processes~\cite{Berkhout-PRL-2008,Ferreira-OL-2011,Saitoh-PRL-2013,Hickmann-PRL-2010,Guo-OL-2009},
there has been few investigations about the propagation property of the optical beam with the OAM but without phase-singularity.
We found that~\cite{Liang-OE-2018} the OAM carried by the EGB with the cross-phase
could result in the AD.
 Specially, when the OAM was equal to a specific value called a critical OAM, the OAM-induced AD could make the elliptic Gaussian mode exist in linearly isotropic media, where only the circular eigenmode is supposed to exist for the beam without the OAM~\cite{Kogelnik-AO-1966,Haus-1984}. At the meantime, the OAM also made the elliptic mode spiral. Our theory can explain why the EGB with the critical OAM can survive in the form of the spiraling elliptic soliton in the media with both linear and nonlinear isotropy, which is ``the long-standing problem'' as pointed out in Ref.~\cite{Desyatnikov-PRL-2010}. Such a spiraling elliptic soliton was predicted first by Desyatnikov et al.~\cite{Desyatnikov-PRL-2010} for local and saturable nonlinearities, and later by us~\cite{Liang-PRA-2013} for the nonlocal nonlinearity with the phenomenological Gaussian response function, but has not been experimentally confirmed thus far.

In this Letter, we experimentally confirm our theoretical prediction about the OAM-induced AD through the propagations of the coherent EGBs with the OAM in the free space and in the cylindrical lead glass, respectively. First, we observe the uniformly-expanding propagation of the spiraling elliptic Gaussian mode in the free space. Second, after showing theoretically that a spiraling elliptic soliton can exist
in the nonlocal nonlinear model with thermal nonlinearity, we observe the stable propagation of such a soliton in the lead glass.
The power-controllable rotation of the EGB is also observed.
The quantitative comparisons with theories are made for both linear and nonlinear cases.

\begin{figure}[tbp]
\centering
\includegraphics[width=70mm]{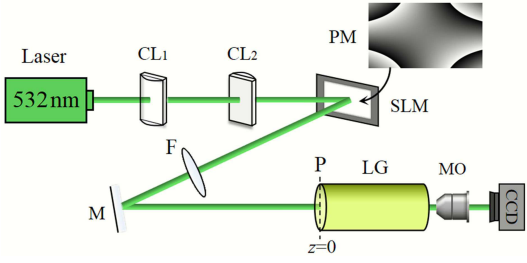}
\caption{Experimental setup. Laser: the green ($\lambda$ = 532 nm) CW laser, CL$_{1,2}$: cylindrical lenses with different focal lengths, SLM: a spatial light modulator, PM: the phase mask, P: the back focal plane of the lens F, LG: a lead glass sample for the nonlinear experiment, MO: a 5$\times$ microscope objective, CCD: a charge-coupled device camera.}
\label{fig1}
\end{figure}

The sketch of our experimental arrangement is shown in Fig.~\ref{fig1}. Two cylindrical lenses CL$_{1}$ and CL$_{2}$ are positioned to transform the circular Gaussian beam emitted from a CW laser into the EGB~\cite{Courtial-OC-1997}. The EGB reflected from the spatial light modulator (SLM) is modulated with a cross phase. The Fourier transform of the EGB is achieved by the lens F at its back focal plane P. For the nonlinear propagation, the front face of the cylindrical lead glass sample is located at the focal plane P, and the output after the sample is recorded by the CCD camera via a microscope objective (MO); while for the linear case in the free space, the beam profiles at different propagation distances are recorded by moving the MO and the CCD camera simultaneously backward from the focal plane P.

In order to carry out the experiment, we first need to determine the OAM carried by the input beam, which can be achieved as following. The phase-modulated EGB reflected from the SLM is of the form
\begin{equation}\label{Gaussian-beam-SLM}
\Phi=
\sqrt{\frac{P}{\pi W_{x}W_{y}}}\exp\left(-\frac{x^{2}}{2W_{x}^{2}}-\frac{y^{2}}{2W_{y}^{2}}\right)\exp(i\Omega xy),
\end{equation}
where $\Phi$ is the electric field scaled by a factor $(2/c\varepsilon_0n_0)^{1/2}$ ($c$ light speed in vacuum, and $n_0$ linear index of the medium) such that the optical power $P=\int|\Phi|^2dxdy$, $W_{x}$ and $W_{y}$ are two semi-axes of the ellipse spot, 
and $\Omega$ ($\Omega>0$ is assumed without loss of generality) is the cross phase coefficient generated by the SLM. We can obtain the input EGB at the focal plane P by the Fourier transform of Eq.~(\ref{Gaussian-beam-SLM})
\begin{equation}\label{input}
\Psi_0
=\sqrt{\frac{P}{\pi w_{0x}w_{0y}}}
\exp\left(-\frac{x^2}{2w_{0x}^2}-\frac{y^2}{2w_{0y}^2}\right)\exp(i\Theta xy),
\end{equation}
where $w_{0x}=\chi/W_{x}$, $w_{0y}=\chi/W_{y}$, $\Theta=W_{x}^{2}W_{y}^{2}\Omega/\chi^{2}$, $\chi=\lambda f\sqrt{1+W_{x}^{2}W_{y}^{2}\Omega^{2}}\big/2\pi$, $\lambda$ the wavelength, and $f$ the focal length of the lens F.
Obviously, $w_{0x}$ and $w_{0y}$ vary with the coefficient $\Omega$. The OAM carried by the EGB above is proportional to $(w_{0x}^{2}-w_{0y}^{2})\Theta$~\cite{Liang-OE-2018,Desyatnikov-PRL-2010}.
It can be deduced that
\begin{equation}\label{Theta}
\Theta=\frac{1}{w_{0x}w_{0y}}\sqrt{\frac{w_{0x}w_{0y}}{w_{\Omega x}w_{\Omega y}}-1},
\end{equation}
where $w_{\Omega x}$ and $w_{\Omega y}$ are the corresponding semi-axes of the input beam when $\Omega=0$. Thus, the OAM per unit input power, $M_0$, can be determined by the following equation (the details are given in Sec.~A of the Supplemental Material~\cite{Supplemental Material})
\begin{equation}\label{OAM}
\frac{M_0}{\Lambda}=\frac{w_{0x}^{2}-w_{0y}^{2}}{2w_{0x}w_{0y}}\sqrt{\frac{w_{0x}w_{0y}}{w_{\Omega x}w_{\Omega y}}-1},
\end{equation}where $\Lambda=1/kc^2$ ($k=2\pi n_0/\lambda$), through the measurement of the input beam widths with and without the cross phase term. Obviously, the sign of the $M_0$ depends on which of $w_{0x}$ and $w_{0y}$ is larger. 

\textbf{Linear Propagations}. As we predicted~\cite{Liang-OE-2018}, the AD could be induced by the OAM carried by the EGB, and the EGB would spiral due to the OAM.
 For such an EGB propagating in linearly isotropic media, the evolutions of its major-axis $w_{maj}~(=w_+)$ and minor-axis $w_{min} ~(=w_-)$ follow~\cite{Liang-OE-2018}
\begin{equation}\label{elliptic-axis}
{w}_{\pm}=\frac{\sqrt{2} w_x w_y}{\sqrt{{w_x^2+w_y^2\mp\sqrt{(w_x^2- w_y^2)^2+w_x^4w_y^4/w_{xy}^2}}}},
\end{equation}
where the functions $w_x(z)$, $w_y(z)$ and $w_{xy}(z)$ are given in Sec.~B of the Supplemental Material~\cite{Supplemental Material}. Here in this paper, we define the elliptic degree
of the beam, $\rho(z)$, by
\begin{equation}\label{ellipticity}
\rho(z)=\frac{w_{maj}(z)}{w_{min}(z)}
\end{equation}
to measure the degree of divergence of ellipses from circles at different distances. Because of its carrying OAM,
the EGB meanwhile spirals with
the rotation angle
\begin{equation}\label{rotion-angle-linear}
\theta=\frac{1}{2} \text{arcsin}\left[\frac{4(M_0/\Lambda)\rho_0^2(1+\rho_0^2)z/z_0}{{\Xi_L}^{1/2}}\right],
\end{equation}
where $\Xi_L=[(1-\rho_0^2)^2(\rho_0^2-z^2/z_0^2)-4(M_0/\Lambda)^2\rho_0^2z^2/z_0^2]^2+[4(M_0/\Lambda)\rho_0^2(1+\rho_0^2)z/z_0]^2$, and $\rho_{0}=\rho(0)~(=w_{0y}/w_{0x})$. As will be seen later, it is set that $w_{0x}<w_{0y}$ in our experiment such that $M_0<0$, thus the EGBs rotate clockwise~\cite{Liang-OE-2018}. While, as expected, EGBs will rotate anticlockwise if $w_{0x}>w_{0y}$. Specially, when the OAM $M_0$ equals positive (or negative) critical OAM~\cite{Liang-OE-2018}
\begin{equation}\label{critical-OAM}
\frac{M_c}{\Lambda}=\frac{(\rho_{0}^{2}-1)^{2}}{4\rho_{0}^{2}},
\end{equation}the elliptic degree $\rho$ will keep unchanged, and the beam evolves in the way of the anticlockwise (or clockwise) spiraling fundamental eigenmode.

\begin{figure}[tbp]
\centering
\includegraphics[width=80mm]{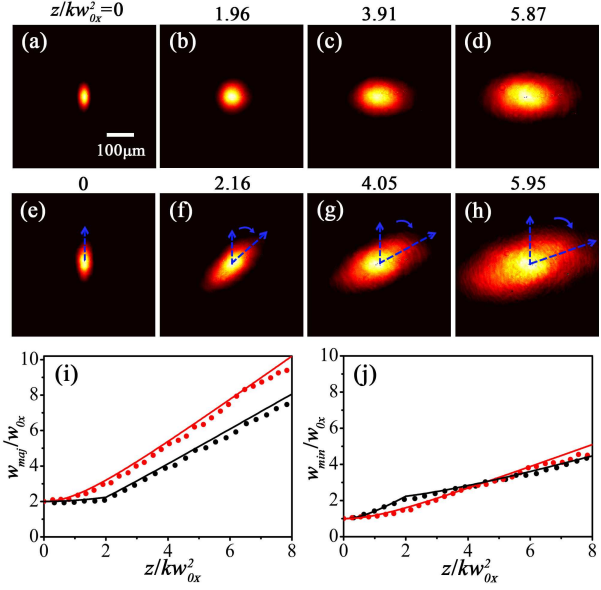}
\caption{Evolutions of the EGBs in the free space with the same $\rho_{0}~(=2.0)$. (a)-(d): $M_0=0$ and $w_{0x}=19.7$~$\mu$m, and (e)-(h): $M_0=-M_c$ and $w_{0x}=25.1$~$\mu$m. (i)-(j): experimental results (solid points) and theoretical results (solid curves) given by Eq.~(\ref{elliptic-axis}) of the semi-axes $w_{maj}$ and $w_{min}$. Black: $M_0=0$, and red: $M_0=-M_c$. }
\label{fig2}
\end{figure}

\begin{figure}[tbp]
\centering
\includegraphics[width=42mm]{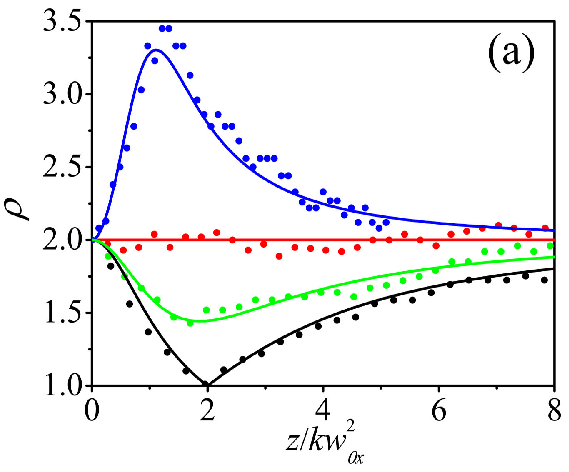}
\includegraphics[width=42mm]{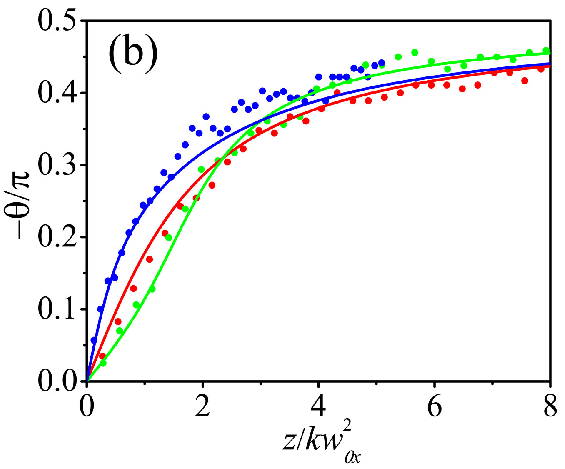}
\caption{Experimental results (solid points) and theoretical results (solid curves), which are given by Eqs.~(\ref{ellipticity}) and (\ref{rotion-angle-linear}) respectively, for 
the elliptic degree (a) and the rotation angle (b) of the EGBs with the same $\rho_{0}~(=2.0)$. Black: $M_{0}=0$, $w_{0x}=19.7$~$\mu$m; green: $M_{0}=-0.30\Lambda$, $w_{0x}=21.2$~$\mu$m; red: $M_{0}=-0.59\Lambda$ (experiment) and $=-0.56\Lambda$ (theory), $w_{0x}=25.1$~$\mu$m (this does be the case with the critical OAM shown in Fig.~\ref{fig2}); blue: $M_{0}=-1.21\Lambda$, $w_{0x}=37.4$ $\mu$m.}
\label{fig3}
\end{figure}

In order to confirm our theoretical prediction above~\cite{Liang-OE-2018}, we experimentally observe, without loss of generality, the evolutions of the input EGBs in the free space with the same initial elliptic degree ($\rho_{0}=2.0$) and the two different OAMs (zero and critical OAMs), as shown in Fig.~\ref{fig2}. Figure~\ref{fig3} presents the comparison between experiments and theories for the elliptic degrees $\rho(z)$ and rotation angles $\theta(z)$ in the same conditions
but four different OAMs. The critical OAM for $\rho_{0}=2.0$ can be determined theoretically by Eq.~(\ref{critical-OAM}) that ${M_c}_T=0.56\Lambda$, and measured experimentally as ${M_c}_E=0.59\Lambda$, which is the input OAM $M_0$ in experiment when the equality $\rho(z)=\rho_0$ almost holds during the entire propagation. For the OAM-free beam in the isotropic linear medium, the beam-expanding by the diffraction is inversely proportional to square of the beam width~\cite{Liang-OE-2018,Chen-OC-2011}. Thus, it can be observed from Figs.~\ref{fig2}(a)-\ref{fig2}(d) that the input OAM-free EGB
[Fig.~\ref{fig2}(a)] spreads more quickly in the $x$-direction than in the $y$-direction. As a result, the elliptic degree $\rho$ decreases initially, and then increases toward its initial value once the elliptic spot evolves from a vertical ellipse into a horizontal one, as shown in Figs.~\ref{fig2}(a)-\ref{fig2}(d) and Fig.~\ref{fig3}(a).
By contrast, the input EGB
carrying the critical OAM [Fig.~\ref{fig2}(e)] expands with the invariant elliptic degree, and rotates clockwise, evolving as the spiraling elliptic Gaussian eigenmode because of the OAM-induced AD~\cite{Liang-OE-2018}, as exhibited in Figs.~\ref{fig2}(e)-\ref{fig2}(h) and also in Fig.~\ref{fig3}(a). The detailed evolutions of two semi-axes for the both input EGBs [Figs.~\ref{fig2}(a) and \ref{fig2}(e)] are shown in Figs.~\ref{fig2}(i) and \ref{fig2}(j), where a well agreement between theory and experiment can be obviously found.
 From Fig.~\ref{fig3}(a) we can also find that the bias of the OAM carried by the beams from the critical OAM makes the elliptic degrees deviate from the initial value during the propagation, and the evolutions of the beams in those cases are not as the fundamental eigenmodes~\cite{Liang-OE-2018}.  It can be concluded that all of the experimental results about the semi-axes [Figs.~\ref{fig2}(i) and \ref{fig2}(j)], the elliptic degree [Fig.~\ref{fig3}(a)] and the rotation angle [Fig.~\ref{fig3}(b)] are consistent with the theoretical results.
\begin{figure}
\centering
\includegraphics[width=45mm]{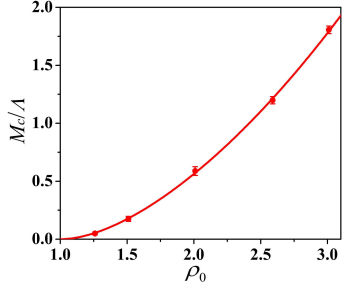}
\caption{Comparison of the experiments (solid points) with the theories (solid curve) given by Eq.~(\ref{critical-OAM}) of the critical OAM for different $\rho_{0}$. }
\label{fig-critical-OAM}
\end{figure}

Furthermore, we measure experimentally the critical OAM ${M_c}_E$ for different initial elliptic degrees, and obtain also the results consistent with the theoretical ones given by Eq.~(\ref{critical-OAM}), as presented in Fig.~\ref{fig-critical-OAM}. We will show later that the relation~(\ref{critical-OAM}), which is the one-to-one dependency of the invariant elliptic degree during the EGB propagation and the OAM carried by the EGB, is a linear property for the propagation of the EGB with the cross-phase, and has nothing to do with the nonlinearity.

\textbf{Nonlinear Propagations}.
Because of the OAM-induced AD, the EGB with the critical OAM will also keep unchanged elliptic degree in the media with both linear and nonlinear isotropy, and spiral at the same time due to the OAM it carried. When the nonlinearity is insufficient, the spiraling EGB will diffract divergently. Otherwise, if the nonlinearity is strong enough to exactly balance the diffraction, the spiraling elliptic soliton will be formed, and both the elliptic degree and the beam width remain unchanged, as has been theoretically predicted~\cite{Desyatnikov-PRL-2010,Liang-PRA-2013}.

Here we report both the theory and the experiment about the nonlinear propagation of the spiraling EGBs in the cylindrical sample of the lead glass. The lead glass itself is of the linear isotropy~\cite{Rotschild-PRL-2005}. On the other hand, its cylindrically symmetric boundary guarantees the nonlinear isotropy~\cite{Shou-OL-2009,Shou-OL-2011,Zhang-OL-2019} because the thermal nonlinearity is of an infinite range of nonlocality and the far-away boundary conditions significantly affect the property of the nonlinear refractive index~\cite{Rotschild-PRL-2005}. Like the saturable nonlinearity, moreover, the thermal nonlinearity is one of the collapse-free nonlinearities, which can preserve the stable (1+2)-dimensional solitons.

The nonlinear propagation of the (1+2)-dimensional optical beam, $\Psi$, in the lead glass can be described by the following coupled equations
\cite{Rotschild-PRL-2005,Shou-OL-2009,Guo-book-2015}
\begin{equation}\label{propagation}
2ik \frac{\partial\Psi}{\partial z}+\nabla^{2}_{\bot}\Psi+2k^2\frac{\Delta n}{n_0}\Psi=0,
\end{equation}
\begin{equation}\label{delta n}
\kappa\nabla^{2}_{\bot}\Delta n=-\alpha\beta|\Psi|^{2},
\end{equation}
where $\alpha$, $\beta$ and $\kappa$ are the absorption, the thermo-optical and the thermal conductivity coefficients.
For the input EGB given by Eq.~(\ref{input}), the rotation due to its carrying OAM is found to be power-dependent (details are provided in Sec.~C of the Supplemental
Material~\cite{Supplemental Material}). Specially when $M_0=\pm M_c$, the rotation angle can be obtained:
\begin{equation}\label{rotation-angle}
\theta(z)=\pm \frac{1}{2}\mathrm{arcsin} \left[\frac{2 \pi  \rho_{0}^{2} (\rho_{0}^{2}+1)\vartheta \sin (\vartheta z/z_0)}{\Xi_{NL}^{1/2}}\right],
\end{equation}
where $\Xi_{NL}=[2 \pi  \rho_{0}^{2} (\rho_{0}^{2}+1)\vartheta \sin (\vartheta z/z_0)]^2+[\alpha_{1}-\alpha_{2}+(\alpha_{1}+\alpha_{2})\cos (\vartheta z/z_0)]^2$, and $\vartheta$, $\alpha_{1}$ and $\alpha_{2}$ are given in the Supplemental Material~\cite{Supplemental Material}.
%
The spiraling elliptic soliton is formed at both the positive (or negative) critical OAM $M_c$ given by Eq.~(\ref{critical-OAM}) and the critical power $P_c$ given by
\begin{equation}\label{powerc}
P_{c}=\frac{2\pi n_{0}\kappa}{\alpha \beta k^{2}w_{0x}^{2}} \frac{(\rho_{0}^{2}+1)^{2}}{4\rho_{0}^{3}},
\end{equation}
and rotates with a constant angular velocity $d\theta/dz=\pm ({\rho_{0}^{2}+1})/2\rho_{0}^{2}/kw_{0x}^{2}$. The elliptic degree of the soliton is determined by the OAM through the OAM-induced AD, i.e., the one-to-one dependency of the invariant elliptic degree and the OAM given by Eq.~(\ref{critical-OAM}), and the power determine only whether the nonlinearity can exactly
balance the diffraction, as the case of the other solitons without the OAM.

\begin{figure}
\centering
\includegraphics[width=70mm]{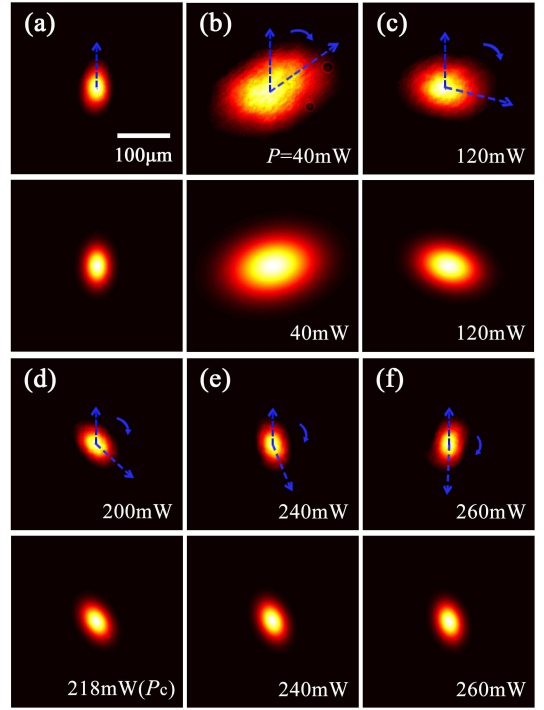}
\caption{Experimental (1st and 3rd row) and theoretical (2nd and 4th row) observations of the spiraling EGBs with $M_0=-M_c$ (${M_c}_T=0.17\Lambda$ and ${M_c}_E=0.16\Lambda$) and $\rho_0=1.50$.
(a): the input beam with $w_{0x}=26.6$~$\mu$m, (b)-(f): the output beams for different input powers, where (d) is the spiraling elliptic soliton.
The experimental and theoretical output widths are, respectively,
(b) $w_{min}=77.4~\mu$m, 65.9~$\mu$m,
(c) $w_{min}=45.0~\mu$m, 34.8~$\mu$m,
(d) $w_{min}=26.4~\mu$m, 26.6~$\mu$m,
(e) $w_{min}=25.0~\mu$m, 26.4~$\mu$m, and
(f) $w_{min}=25.5~\mu$m, 26.4~$\mu$m.
The experimental output elliptic degree $\rho_{out}=1.43,~1.41,~1.47,~1.48$, and 1.54, respectively. The theoretical output intensities are obtained from the analytic solutions to the Snyder-Mitchell model [see Eq.(C.6) in Supplemental Material].}
\label{fig4}
\end{figure}

In order to demonstrate our theoretical results above, we carried out the experiments for the nonlinear dynamics of the EGBs with the OAM in the cylindrical sample of the lead glass, which was used in our early experiments~\cite{Shou-OL-2009,Shou-OL-2011,Zhang-OL-2019}, with the length $L=57.5$mm and  the radius $R=7.5$mm. The other parameters of the sample are $n_{0}=1.9$, $\alpha=0.06$cm$^{-1}$, $\beta=14\times 10^{-6}$K$^{-1}$ and $\kappa=0.7$W(m$\cdot$ K)$^{-1}$~\cite{Zhang-OL-2019}. The experimental setup is shown in Fig.~\ref{fig1}.

Figure~\ref{fig4} gives the intensity distributions of the input beam [Fig.~\ref{fig4}(a)]
 and the output beams [Figs.~\ref{fig4}(b)-\ref{fig4}(f)] after propagating the distance $L/k w_{0x}^2=3.6$ for different input powers. At relatively low power, i.e., $P=40$mW, the self-focusing is too weak to overcome diffraction, thus the output beam has a significantly larger spot and rotates clockwise at the same time due to its carrying OAM [Fig. \ref{fig4}(b)]. As the power increases, the EGB undergoes stronger self-focusing so that its spot shrinks gradually and rotates more rapidly at the meantime, until the diffraction is exactly balanced by the self-focusing to form an elliptic soliton with the measured values $w_{min}=26.4$$\mu$m, $\rho_{out}=1.47$, $P_c=200$mW and $\theta=-145^{\circ}$ [Fig.~\ref{fig4}(d)],
as compared to the theoretical values $w_{min}=26.6$~$\mu$m, $\rho_{out}=1.50$, $P_c=218$mW and $\theta=-149^{\circ}$, respectively.
A further increase of the power leads to the smaller beam spot than the input beam, and a monotonous increase of the rotation angle. The difference from what we are observing here is that the elliptic solitons without the OAM do not spiral, as has been experimentally observed~\cite{Rotschild-PRL-2005,Katz-OL-2004,Zhang-OE-2007}.

\begin{figure}[htbp]
\centering
\includegraphics[width=42mm]{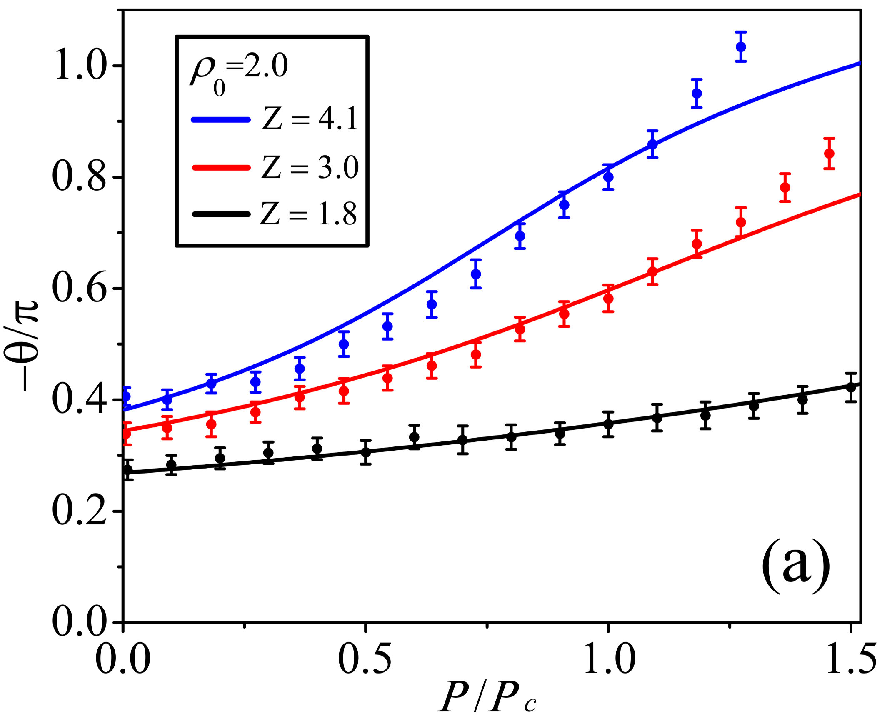}
\includegraphics[width=42mm]{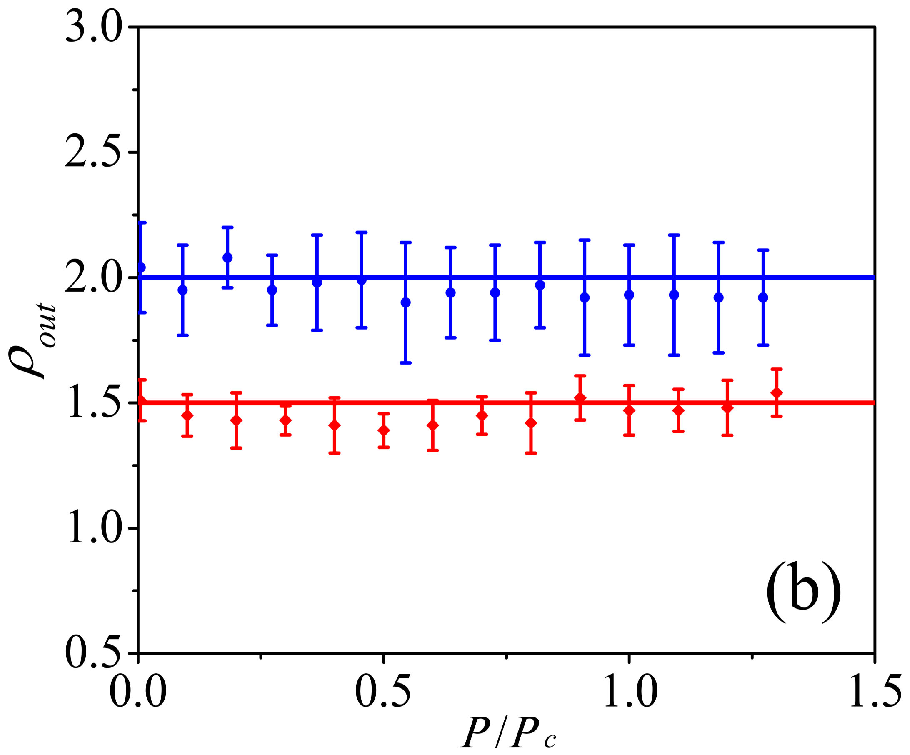}
 \caption{(a) Experimental results (solid points) and theoretical results (solid curves) given by Eq.~(\ref{rotation-angle}) of the rotation angles.
  Black: $w_{0x}=37.5$$\mu$m, red: $w_{0x}=29.2$$\mu$m, blue: $w_{0x}=25.1$$\mu$m. $Z=L/k w_{0x}^2$, and $\triangle P$=10mW, 15mW and 20mW for black, red and blue cases, respectively. (b) Experimental data of the output elliptic degree $\rho_{out}$ for the EGBs with the critical OAM but different powers. Blue: $\rho_{0}=2.0$, $L/kw_{0x}^2=4.1$; red: $\rho_{0}=1.5$, $L/kw_{0x}^2=3.6$. The straight lines are the corresponding theoretical ones. For both of figures, the abscissa axis is normalized by the corresponding theoretical critical powers.}
\label{fig5}
\end{figure}
To further confirm the nonlinear dynamical properties of the spiraling elliptic breathers~\cite{Liang-OE-2015} and solitons~\cite{Liang-PRA-2013,Desyatnikov-PRL-2010}, we carried out a series of experiments for input beams with different sizes and different input powers in the case $M_0=-M_c$. Figure~\ref{fig5}~(a) is the rotation angle at the sample output as a function of the input power for the input beams with $\rho_{0}=2.0$
but different $w_{0x}$. 
The power intervals for the measuring, $\triangle P$, are chosen such that the ratios of $\triangle P/P_c$ are
almost equal. We can observe that the rotation angles of the EGBs are mainly in agreement with the theoretical results whatever the nonlinear case [Fig.~\ref{fig5}~(a)] or linear case [Fig.~\ref{fig3}~(b)]. Moreover, the rotation angle can exceed $\pi/2$ in the former, and cannot in the latter.
 It is also observed from Fig.~\ref{fig5}~(a) that the power-controllable rotation of the EGBs can be realized, with a rotation angle reach to 190$^{\circ}$ over about 4 diffraction lengths. We also measured the critical powers of the solitons for different beam parameters. The experimental results are presented in Table~\ref{critical-power}, where the theoretical ones based on Eq.~(\ref{powerc}) are also given. Obviously, the theory is in very close agreement with the experiment with the maximum absolute value of relative errors not more than 10 percent.
\begin{table}[tbp]
\renewcommand\arraystretch{1.2}
\caption{The critical powers $P_c$ for the solitons.}\label{critical-power}
\center
\begin{tabular}{p{1.7cm}<{\centering}| p{0.95cm}<{\centering} p{0.95cm}<{\centering} p{0.95cm}<{\centering} p{0.95cm}<{\centering} p{0.95cm}<{\centering} p{0.95cm}<{\centering}}
\hline\hline
$\rho_0$ & 1.5 & 1.5 & 1.5 & 2.0 & 2.0 & 2.0\\
\hline
$w_{0x}$($\mu$m) & 42.0 & 31.5 & 26.6 & 37.5 & 29.2 & 25.1\\
\hline
E (mW)\footnote{E: experiment, T: theory, RE: relative error, RE=(E$-$T)/T.} & 90 & 160 & 200 & 100 & 165 & 220\\
T (mW) & 88 & 156 & 218 & 110 & 181 & 245\\
RE ($\%$) & $2.3$ & $2.6$ & $-8.3$ & $-9.1$ & $-8.8$ & $-10$\\
\hline\hline
\end{tabular}
\\[0.1cm]
\end{table}

In the end of the nonlinear part, we show the fact that the OAM-induced AD is a linear effect and has nothing to do with nonlinearity, as can be observed in Fig.~\ref{fig5}~(b), which presents that both the elliptic breathers and the elliptic solitons preserve their initial elliptic degree $\rho_0$.

In conclusion, through both theory and experiment, we discover a novel phenomenon that the OAM carried by the EGB can induce the AD.
The OAM-induced AD can make the EGB with critical OAM evolve as the spiraling elliptic Gaussian (fundamental) eigenmode, which expands with the invariant elliptic degree, in the linearly isotropic medium. At both
the critical OAM and the critical power, the spiraling elliptic soliton
can be formed and stably propagate in the medium with both the isotropic linearity and the isotropic collapse-free nonlinearity.
The elliptic degree of the soliton is determined by the OAM through the OAM-induced AD, which is a linear effect.
 The power determine only whether the nonlinearity can exactly
balance the diffraction.
The rotation of the EGBs due to their carrying OAM is power-controllable.
We expect that such power-controllable spiraling EGBs may have applications to optical spanners in bio-photonics, opto-fluidics, and life sciences.



\begin{thebibliography}{99}

\bibitem{Born-book}M. Born and E. Wolf, \textit{Principles of Optics} (Pergamon Press, Inc., New York, 1980), Chap. 8, 6th ed.

\bibitem{Fleck-josa-1983}J. A. Fleck and M. D. Feit, J. Opt. Soc. Am. \textbf{73}, 920 (1983).

\bibitem{Ciattoni-pre-2002}A.Ciattoni, G. Cincotti, D. Provenziani, and C. Palma,
Phy. Rev. E \textbf{66}, 036614 (2002).

\bibitem{Chen-OC-2011}
Z. Chen and Q. Guo, Opt. Commun. \textbf{284}, 183 (2011).


\bibitem{Guo-JOA-2000}
Q. Guo and C. Sien, J. Opt. A: Pure Appl. Opt. \textbf{2}, 5 (2000).


\bibitem{Liang-OE-2018}
G. Liang, Y. Wang, Q. Guo, and H. Zhang, Opt. Express \textbf{26}, 8084 (2018).

\bibitem{Polyakov-PRE-2002}
S. V. Polyakov and G. I. Stegeman, Phy. Rev. E \textbf{66}, 046622 (2002).

\bibitem{Kogelnik-AO-1966}
H. Kogelnik and T. Li, Appl. Opt. \textbf{5}, 1550 (1966).

\bibitem{Haus-1984}
H. A. Haus, \textit{Waves and Fields in Optoelectronics}, (Prentice-Hall, 1984), Chapt. 4 and 5.

\bibitem{Seshadri-JOSAA-2001} S. R. Seshadri, J. Opt. Soc. Am. A \textbf{18}, 2628  (2001).

\bibitem{Seshadri-JOSAA-2003}S. R. Seshadri, J. Opt. Soc. Am. A \textbf{20}, 1818 (2003).

\bibitem{Ciattoni-oc-2004} A. Ciattoni, and C. Palma,  Opt. Commun., \textbf{231}, 79 (2004).

\bibitem{Crosignani-OL-1993}
B. Crosignani and P. Di Porto, Opt. Lett. \textbf{18}, 1394 (1993).

\bibitem{Snyder-OL-1997}
A. W. Snyder and D. J. Mitchell, Opt. Lett. \textbf{22}, 16 (1997).

\bibitem{Tichonenko-OL-1998}
V. Tichonenko, Opt. Lett. \textbf{23}, 594 (1998).

\bibitem{Eugenieva-OL-2000}E. D. Eugenieva, D. N. Christodoulides, and M. Segev, Opt. Lett. \textbf{25}, 972 (2000).

\bibitem{Katz-OL-2004}O. Katz, T. Carmon, T. Schwartz, M. Segev, and D. N. Christodoulides, Opt. Lett., \textbf{29}, 1248 (2004).

\bibitem{Conti-PRE-2005}
C. Conti, M. Peccianti, and G. Assanto, Phy. Rev. E \textbf{72}, 066614 (2005).


\bibitem{Zhang-OE-2007}
P. Zhang, J. Zhao, C. Lou, X. Tan, Y. Gao, Q. Liu, D. Yang, J. Xu, and Z. Chen,
Opt. Express \textbf{15}, 536 (2007); P. Zhang, J. Zhao, F. Xiao, C. Lou, J. Xu, and Z. Chen, \emph{ibid.} \textbf{16}, 3865 (2008).



\bibitem{Rotschild-PRL-2005}
C. Rotschild, O. Cohen, O. Manela, M. Segev, and T. Carmon, Phys. Rev. Lett. \textbf{95}, 213904 (2005).


\bibitem{Christodoulides-PRL-1998}
D. N. Christodoulides, T. H. Coskun, M. Mitchell, and M. Segev, Phys. Rev. Lett. \textbf{80}, 2310 (1998).

\bibitem{Allen-pra-1992}
L. Allen, M. W. Beijersbergen, R. J. C. Spreeuw, and J. P. Woerdman,
Phys. Rev. A \textbf{45}, 8185 (1992).

\bibitem{Grier-Nature-2003}
D. G. Grier, Nature \textbf{424}, 810 (2003).

\bibitem{Dholakia-NP-2011}
K. Dholakia and T. Cizmar, Nat. Photonics \textbf{5}, 335 (2011).

\bibitem{Volke-Sepulveda-job-2002}
K. Volke-Sepulveda, V. Garc\'{e}s-Ch\'{e}z, S. Ch\'{a}v{e}z-Cerda, J. Arlt, and K. Dholakia,
J. Opt. B \textbf{4}, S82 (2002).

\bibitem{Gutierrez-Vega-proc-spie-2008}
J. C. Guti\'{e}rrez-Vega, 
Proc. of SPIE, \textbf{7062}, 706207 (2008).

\bibitem{Desyatnikov-PRL-2010}
A. S. Desyatnikov, D. Buccoliero, M. R. Dennis, and Y. S. Kivshar, Phys. Rev. Lett. \textbf{104}, 053902 (2010).

\bibitem{Liang-PRA-2013}
G. Liang and Q. Guo, Phys. Rev. A \textbf{88}, 043825 (2013).

\bibitem{Allen-book-1999}L. Allen, M. J. Padgett, and M. Babiker, Prog. Opt. \textbf{39}, 291 (1999).

\bibitem{Courtial-OC-1997}
J. Courtial, K. Dholakia, L. Allen, and M. J. Padgett, Opt. Commun. \textbf{144}, 210 (1997).

\bibitem{Kotlyar-OE-2018}
V. V. Kotlyar, A. A. Kovalev, and A. P. Porfirev, Opt. Express \textbf{26}, 141 (2018).


\bibitem{Berkhout-PRL-2008}
G. C. G. Berkhout and M. W. Beijersbergen, Phys. Rev. Lett. \textbf{101}, 100801 (2008).

\bibitem{Hickmann-PRL-2010}
J. M. Hickmann, E. J. S. Fonseca, W. C. Soares, and S. Ch\'{a}vez-Cerda, Phys. Rev. Lett. \textbf{105}, 053904 (2010).

\bibitem{Saitoh-PRL-2013}
K. Saitoh, Y. Hasegawa, K. Hirakawa, N. Tanaka, and M. Uchida, Phys. Rev. Lett. \textbf{111}, 074801 (2013).

\bibitem{Guo-OL-2009}
C. Guo, L. Lu, and H. Wang, Opt. Lett. \textbf{34}, 3686 (2009).

\bibitem{Ferreira-OL-2011}
Q. S. Ferreira, A. J. Jesus-Silva, E. J. S. Fonseca, and J. M. Hickmann, Opt. Lett. \textbf{36}, 3106 (2011).

\bibitem{Supplemental Material} See Supplemental Material at http:  for detailed derivation process
and other relevant information.


\bibitem{Shou-OL-2009}
Q. Shou, Y. Liang, Q. Jiang, Y. Zheng, S. Lan, W. Hu, and Q. Guo, Opt. Lett. \textbf{34}, 3523 (2009).

\bibitem{Shou-OL-2011}
Q. Shou, X. Zhang, W. Hu, and Q. Guo, Opt. Lett. \textbf{36}, 4194 (2011).


 \bibitem{Zhang-OL-2019}
H. Zhang, M. Chen, L. Yang, B. Tian, C. Chen, Q. Guo, Q. Shou, and W. Hu, Opt. Lett. \textbf{44}, 3098 (2019).

%
%
%
%
%
%


%


\bibitem{Guo-book-2015}
Q. Guo, D. Lu, and D. Deng, in \emph{Advances in Nonlinear Optics}, edited by X. Chen, Q. Guo, W. She, H. Zeng,
and G. Zhang, (De Gruyter, 2015), Chap. 4, pp. 227-305.

\bibitem{Liang-OE-2015}
G. Liang, Q. Guo, W. Cheng, N. Yin, P. Wu, and H. Cao, Opt. Express \textbf{23}, 24612 (2015).

\end{thebibliography}
\end{document}